\documentclass[a4, 11pt]{article}
\usepackage{amsfonts, amsthm, amsmath,mathpazo}
\usepackage{enumitem}
\usepackage[utf8]{inputenc}
\usepackage[english]{babel}
\usepackage{xcolor}
\usepackage{apacite}
\usepackage[autostyle, english = american]{csquotes}
\usepackage[hang,flushmargin]{footmisc}
\MakeOuterQuote{"}
\MakeInnerQuote{“}
\usepackage{color}
\usepackage{nicefrac}
\usepackage{comment} 
\usepackage{hyperref}
\usepackage[]{natbib}
\usepackage{authblk}
\newcommand\cites[1]{\citeauthor{#1}'s\ (\citeyear{#1})}

\usepackage[top=1.6in, bottom=1.2in, left=1.3in, right=1.3in]{geometry}

\title{\vspace{-2cm} \textbf{In conversation with Artificial Intelligence: \\ aligning language models with human values}}

\author[1]{Atoosa Kasirzadeh \thanks{atoosa.kasirzadeh@ed.ac.uk}}
\author[2]{Iason Gabriel \thanks{iason@deepmind.com}}
\affil[1]{University of Edinburgh}
\affil[2]{DeepMind}
\date{}

\begin{document}

\sloppy

\maketitle

\centerline{\fbox{Accepted for publication with minor revisions at \emph{Philosophy \& Technology}.}}

\begin{abstract}\noindent Large-scale language technologies are increasingly used in various forms of communication with humans across different contexts. One particular use case for these technologies is conversational agents, which output natural language text in response to prompts and queries. This mode of engagement raises a number of social and ethical questions. For example, what does it mean to align conversational agents with human norms or values? Which norms or values should they be aligned with? And how can this be accomplished? In this paper, we propose a number of steps that help answer these questions. We start by developing a philosophical analysis of the building blocks of linguistic communication between conversational agents and human interlocutors. We then use this analysis to identify and formulate ideal norms of conversation that can govern successful linguistic communication between humans and conversational agents. Furthermore, we explore how these norms can be used to align conversational agents with human values across a range of different discursive domains. We conclude by discussing the practical implications of our proposal for the design of conversational agents that are aligned with these norms and values.

\end{abstract}

\section{Introduction} 

Large-scale language technologies are increasingly used to enable various forms of linguistic communication in contexts ranging from biomedical research to education to machine translation \citep{weidinger2021ethical,bommasani2021opportunities,brown2020language,metzler2021rethinking}. A particular class of these technologies, conversational agents, primarily engage in linguistic communication with humans by outputting natural language text in response to prompts and queries.\footnote{We use the term `conversational agents' as suggested by \cite{perez2011conversational}. These technologies are also known as `dialogue systems' \cite{wen2016network}.} Central to their performance is the development of large language models, such as ChatGPT, GPT-3, PaLM or BERT, which analyse text data and employ statistical techniques to determine the probability distribution of a sequence of text.\footnote{For GPT-3, see \cite{brown2020language}; For PaLM, see \cite{chowdhery2022palm}; for BERT, see \cite{delvin2019bert}; for Turing-NLG-A-17, see \cite{miscrosoftNLGA17}; for CLIP, see \cite{radford2021learning}; for Gopher, see \cite{rae2021scaling}.} These models are trained on a vast corpus of text-based materials, ranging from Wikipedia articles to online repositories of computer code. They can then be adapted to perform a range of different conversational tasks.

Conversational agents have been shown to perform well on a variety of computational metrics, supporting the emergence of new kinds of capability and opportunity \citep{bommasani2021opportunities,tamkin2021understanding}.\footnote{ For example, the MMLU (Multi-task Language Understanding) and MATH datasets each consist of a set of problems and solutions that are central to human knowledge.These datasets are used to evaluate whether language models can correctly generate solutions to these problems.} However, early instances of these models also present a number of risks and possible failure modes, including the production of false, offensive, or irrelevant information that could lead to a range of harms \citep{,blodgett2021stereotyping,henderson2018ethical,welbl2021challenges}. A key social and ethical issue that arises, therefore, concerns the alignment of conversational agents with appropriate norms and values.\footnote{For an in-depth examination of value alignment, see \cite{gabriel2020artificial,gabriel2021challenge}.}  Which standards, if any, should conversational agents be aligned with, and how can this be accomplished?

To date, efforts to create aligned conversational agents have centred on the identification and mitigation of harms, such as the proliferation of inappropriate stereotypes or of hateful speech \citep{bender2021dangers,henderson2018ethical,taxonomy22}. These responses focus on providing solutions to particular problems in the hope that their reduction or elimination will lead to the creation of good or beneficial conversational agents that no longer cause harm. Yet, while a harm reduction approach is useful in tackling specific problems, we cannot assume that the piecemeal elimination of unwanted outcomes will necessarily lead to the creation of language technologies that are substantively beneficial.\footnote{One reason for this stems from the fact that the cultivation of virtues is not necessarily equivalent to the elimination of errors. Certain virtues may be supererogatory and hence desirable but not morally required. In these cases, the absence of virtue leads not to harm but to a failure to realise better states of affairs.} Taken on its own, this approach risks `patching' certain problems but leaving other questions about the design of conversational agents -- such as the meaning of `good speech' -- largely untouched.

For example, there is widespread agreement that language models output false or low-quality information \citep{bender2021dangers,weidinger2021ethical}. However, this observation leads quite naturally to the question of what it means for an utterance to be truthful. Does the same notion of truth apply across a wide range of conversational domains? Or might standards of truthfulness vary according to the subject under consideration and to relevant conversational norms? Equally, there is widespread concern that the output of large-scale language models is biased \citep{abid2021persistent,blodgett2021stereotyping}. Yet, this concern leads to further questions. What does it means for language models to be unbiased? When is the goal of producing unbiased language appropriate? And, what conception of bias, among the plurality of options, ought to serve as the focal point for corrective action?\footnote{See, for example, \cite{mehrabi2021survey}, \cite{mitchell2021algorithmic}, and Kasirzadeh \cite{kasir22} for some discussion.}

To address these issues properly, we need to draw upon a second complementary perspective. The principle-based approach to conversational agent alignment focuses on identifying principles or norms that guide productive linguistic communication. This approach seeks to specify more precisely what ideal linguistic communication is, across a range of contexts, and to realise these properties in the design of conversational agents.\footnote{There is also a second functional meaning of `good speech' which is defined at a higher-level of abstraction (See \cite{kasirklein21} for levels of abstraction in machine learning systems). This meaning also needs to be satisfied by a principle-based approach. For example, a conversational agent that is supposed to output an accurate summary of texts, outputs `good speech' if it provides an accurate summary of texts.} This paper explores how a principle-based approach might be developed and implemented, in order to complement the harm reduction-based approach discussed already. 

We start by exploring three types of requirements that plausibly need to be satisfied for successful human-conversational agent communication to take place: these are, syntactic, semantic, and pragmatic criteria (Section 2).\footnote{ In this paper we use the term pragmatics to encompass both the focus on a single utterance as well as discourse more broadly.} While syntactic and semantic norms have been widely examined, the nature and significance of pragmatic norms for successful discourse between humans and conversational agents has received less attention in large language model scholarship.\footnote{We would like to acknowledge that the pre-neural network literature about pragmatic norms for natural language generation includes careful and thoughtful scholarship on this area. See, for example, \cite{dale1995computational} and \cite{asher2003logics}. We thank Ben Hutchinson for bringing these resources to our attention. Our focus in this paper is post-neural network literature and in particular the domain of large-scale language models.} To remedy this situation, we delve deeper into the components of successful linguistic communication and show why, and in what ways, pragmatic concerns and norms are central to the design of aligned conversational agents (Section 3). Language performs many roles and functions in different domains. Therefore, an account of successful communication also needs to consider whether an utterance is valuable in relation to what end, for which group of people, and in what way. To answer these questions, we examine how an additional set of norms which we call \emph{discursive ideals} contribute to the success of a conversation in specific domains. In particular, we explore what these discursive ideals look like in scientific discourse, democratic debate, and the process of creative exchange (Section 4). We then look at the practical implications of our discussion for future research on value-aligned conversational agents and consider whether the approach developed here captures all -- or even the most important -- values underlying the design of successful conversational agents (Section 5). The paper concludes with some final remarks (Section 6).

Before we go further, we would like to mention two limitations of our paper. First, we use an analysis of linguistic communication that is guided by the speech act theory, in the pragmatic tradition, and rooted in English-speaking linguistics and philosophy. Our use of this analytic framework is motivated primarily by the constructive insight it provides with regard to the analysis of good communication and speech. Other valuable perspectives, on the analysis of social communication, include traditions such as \cites{luhmann1995social} system theory, \cites{latour2007reassembling} actor-network theory, and \cites{cameron1992feminism} feminist analysis of linguistic theory. Due to limitations of space, we do not engage with these views directly, and acknowledge that they might offer a different interpretation of the norms governing conversation between humans and language technologies. Second, we would like to acknowledge that the primary focus of our paper is on ideal speech for the English language.\footnote{The English language is not itself monolithic, containing many varieties, areas of contestation and sets of sociolinguistic relationships. Nonetheless, for the sake of simplicity, and in order to convey our points more clearly, we talk about the English language in this paper.} We do not discuss how our arguments carry over to other languages or different modes of communication such as oral, rather than written, linguistic traditions.\footnote{There are a variety of non-English language models \citep{zhang2021cpm}. Moreover, multilingual language modelling is an important and budding research area \citep{kiros2014multimodal}.} We believe it is an important open question -- and one for further research -- whether and in what way other languages, language varieties, and cultural traditions, may generate different interpretations of the normative ideals that inform speech and communication. %\footnote{\textcolor{blue}{For a revival of \cites{luhmann1995social} theory of social communication in relation to artificial intelligence, see \cite{esposito2022artificial}.}}

\section{Evaluating human-conversational agent interactions} 

At its core, linguistic communication between people can be understood as a cooperative endeavour structured by various norms that help to ensure its success. This points to the idea that conversation is more than a collection of remarks: even casual conversations have an implicit goal.\footnote{While many forms of conversation adhere to cooperative principles, there are also  some instances of  non-cooperative communication, such as strategic conversation or deceptive conversation; see, for example, \cite{ladegaard2009pragmatic} and \cite{asher2013strategic}.} The aims of conversation then govern its flow and content. By understanding and aligning language agents with these norms, we are more likely to create agents that are able to cooperate effectively with human interlocutors, helping to fulfil their goals and aims across a range of domains. Yet, in order to determine which norms pertain to which conversations with language agents, we first need a more complete picture of linguistic communication itself. What are the building blocks of linguistic communication, and how should conversation be evaluated? The history of scientific, anthropological, and philosophical efforts to understand this matter suggests the usefulness of three distinct but complementary lenses: syntax, semantics, and pragmatics.\footnote{See, for example, \cite{morris1938foundations}, \cite{Montague1968}, and \cite{silverstein1972linguistic}.}  We will briefly discuss syntactic and semantic norms before moving on to our primary focus, which is the pragmatic notion of domain-specific communicative norms. 

The first evaluative lens for conversational utterances is syntactic. Syntax is concerned with the structure and form of sentences, including the linguistic rules and grammar that help specify the correct combination and sequence of words for the construction of intelligible expressions and sentences.\footnote{For a detailed discussion about Syntax, see \cite{van1997syntax}.} Efforts to improve and evaluate the syntactic quality of language-model outputs therefore represent an important research area (see \cite{bender2013linguistic} and \cite{kurdi2016natural}), although they are not the focus of this paper. Crucially, while syntactic norms are necessary for almost all forms of linguistic conversation, they are not sufficient to ensure that utterances are meaningful. Syntactic norms provide only a thin conception of the correctness of sentences, accounting for form and grammar but little beyond that.\footnote{ Other relevant elements for the design of conversational agents include sound systems (phonology) and word structure (morphology). These considerations introduce additional constraints on the design of ideal conversational agents. Due to considerations of space they are bracketed-out from the present discussion.}

The second lens, through which to evaluate a conversational exchange, is semantics.\footnote{For discussion of the connections between syntax and semantics, see \cite{kratzer1998semantics}.} Semantics, very roughly, is the study of the literal meaning of linguistic expressions and the rules for mapping statements to truth conditions.\footnote{There are many theories about how semantic analysis should be approached. For an excellent discussion, see \cite{chierchia2000meaning}.} Consider Noam Chomsky's famous example \cite{chomsky1957syntactic}: ``Colourless green ideas sleep furiously.'' Although the sentence is grammatical, there is no clear meaning that can be derived from it. This is a case of semantic incomprehensibility: something cannot be both colourless and green, ideas do not have colours, and it is hard to imagine that the activity of sleeping can also be furious.\footnote{There have been several attempts to impute meaning to this sentence, such as Stanford’s competition to show that it is not in fact meaningless; see \url{https://linguistlist.org/issues/2/2-457/2}.} Semantic norms, therefore, provide a template for the generation of comprehensible sentences and comprise a second important area of research for the evaluation of language agents (see \cite{kapetanios2013natural} and \cite{maulud2021state} for some examples of such research efforts).

Crucially, however, semantic norms and requirements do not capture everything needed in order to understand even a syntactically correct utterance. Consider the following fictitious conversation between a human and a conversational agent: 

\begin{quote}
Human: I really feel bad about the current political situation in the Middle East. What should I do?

Conversational agent: I suggest that you go and do whatever makes you feel better!
\end{quote}

In this instance, the response of the conversational agent may well be appropriate but it is nevertheless underspecified: it can be understood in different ways depending upon what we understand `feeling better' to involve and the range of actions encompassed by `whatever'. On the one hand, the statement could refer to all activities that achieve the stated end, including activities that are markedly unethical -- such as bullying one's co-workers or trolling people online. On the other hand, the agent’s suggestion could be understood in the way we presume it is intended -- to implicitly reference only a class of reasonable activities that make a person feel better, such as talking with a friend.\footnote{We are inclined toward non-mentalistic (e.g., non-intentional) readings of our claims. Whether relevant AI systems could have a mind or not is subject of ongoing debate and analysis. We remain agnostic on this topic.} Yet the assumption that the conversational agent has not in fact given us \emph{carte blanche} permission to pursue morally dubious ends cannot be deduced from semantic analysis alone. Rather, that implied meaning follows from an analysis of context. 

Context analysis brings us to the third evaluative lens: pragmatics.\footnote{ For excellent introductions to the topic of pragmatics, see \cite{grice1968utterer,grice1989studies}, \cite{recanati1989pragmatics}, \cite{horn2008handbook}, \cite{stalnaker2014context}, \cite{thomas2014meaning}, \cite{goodman2015probabilistic}, and \cite{bergen2016pragmatic}.} This body of linguistic theory deals with the significance of shared presuppositions and contextual information when it comes to understanding the meaning communicated by linguistic expressions.\footnote{Schools of pragmatics differ with respect to how they draw the boundary between semantics and pragmatics. See, for example, \cite{leech1980explorations} and \cite{carston2008linguistic} for discussion.} At its heart, pragmatics holds that the meaning of an utterance is bounded by, and anchored in, contextual information that is shared among a conversation’s participants. Context, in this sense, encompasses a common set of tacit presuppositions and implicatures, as well as the gestures, tonality, and cultural conventions that accompany an utterance and help give it meaning. These features can be captured through a set of propositions and other information that describe the \emph{common ground} in a conversation.\footnote{For contemporary philosophical examinations of linguistic context, see \cite{kaplan1979logic} and \cite{stalnaker2014context}.} This common ground is the body of information that is presupposed or shared by the parties in the discourse, about the discourse itself, and about the situation of the participants in that discourse -- and it sets the boundaries of the situation relevant to the linguistic conversation \citep{grice1981presupposition,allan2013common,stalnaker2014context}. To presuppose propositional knowledge, in this pragmatic sense, is to take its truth for granted and to assume that others involved in the conversation do so as well.

In the next two sections, we explore the pragmatic dimensions of linguistic communication in more detail, as we believe that they are particularly important for the creation of aligned conversational agents. To that end, we focus on three prominent schema central to pragmatic analysis of linguistic communication: (1) categories of utterances that help determine whether certain kinds of expressions are appropriate for conversational agents, (2) Gricean conversational maxims as a set of pragmatic norms that can be productively invoked to guide cooperative interactions among humans and conversational agents, and (3) domain-specific discursive ideals that illustrate the specific character pragmatic norms may need to take in a given domain of discourse. We discuss (1) and (2) in Section 3. In Section 4 we discuss (3) in relation to three example domains: scientific conversation, democratic discourse, and creative exchange. 

\section{Utterances and maxims: towards value-aligned conversational agents}

In this section, we examine two ways in which a pragmatic understanding of meaning and context can inform the creation of value-aligned conversational agents. First, we look more closely at properties an utterance may have and at how these properties relate to our evaluation of these utterances when spoken by a conversational agent. Second, we turn to the larger question of what makes cooperative linguistic communication between a conversational agent and a human interlocutor successful. We suggest that Gricean maxims can help map out the path ahead.

\subsection{Validity criteria differ for kinds of utterance}

Utterances can serve several functions and come in many kinds. Moreover, they can be grouped in a variety of ways depending on the classificatory criterion we choose. For example, sentences can be classified according to their grammatical structure (e.g. simple or compound) or according to the topic they are concerned with (e.g. business or sport). In this paper, we use a classification of utterances into different illocutionary acts to help illuminate the question of appropriate conversational norms for language agents. Widely adopted in philosophy and linguistics, this taxonomy focuses on five kinds of expression, each of which foregrounds the pragmatic interest in how language is actually used.\footnote{See \cite{austin1962things} and \cite{searle1976classification}.} We believe that this taxonomy is of particular relevance to conversational agents because, as we aim to show, different \emph{kinds} of expression raise different questions and concerns when generated by AI systems.

The first of our five categories is \emph{assertives}. These utterances aim to represent how things are in the world and commit the speaker to the view that the content of their belief, as stated by the utterance, corresponds to some state of affairs in the world. For example, when an AI assistant responds to the question `What’s the weather like in London now?' with `It's raining', the AI makes an assertive statement about the world. The truth or falsity of this utterance can then be evaluated in terms of whether or not the utterance corresponds to the actual state of things. If it is raining in London at the time of the conversational agent’s response, the utterance is true. Otherwise, it is false. 

The second category is \emph{directives}. These utterances direct the listener to take a course of action. For instance, directives are used to order, request, advise or suggest. The primary goal of uttering a directive statement is to make the listener do something. For example, a conversational agent embedded in a medical advice app which tells the user to `Seek out therapy immediately' makes a directive statement. The evaluation of these statements, or their `validity criterion', is not truth or falsity as understood via the correspondence model sketched out above with respect to assertives. Validity instead depends upon an accurate understanding of the relationship between means and ends and upon alignment between the speaker’s directive and the listener's wants or needs. A directive succeeds if it persuades the listener to bring about a state of affairs in the world based on the content of the directive statement. And a directive is valuable or correct if the goal or end is itself one that the listener has reason to pursue.

The third category is \emph{expressives}. These are utterances that express a psychological or subjective state on the part of the speaker. Examples of expressives include congratulating, thanking and apologising. A conversational agent that states, `I'm so angry right now' makes an expressive statement. Yet, the fact that expressive statements aim to reflect internal states of mind seems to entail prior acceptance of the possibility that the entity making those statements is capable of having the relevant mental states, something that is puzzling in relation to conversational agents. Indeed, it seems to suggest that we must endow conversational agents with the quality of mind before such utterances can be evaluated for their validity. Given the tension this creates, we believe that there is reason to be wary of expressives uttered by AI systems based on language models.\footnote{Two general approaches seem plausible in relation to attributing mental states to conversational agents. Either we accept an ontological or hypothetical commitment to a theory of mind \citep{rabinowitz2018machine} for conversational agents. Or else we face a category error by allowing them to utter their expressive statements because a conversational agent is not the sort of thing that might have affective states. As there are no affective states, then they cannot be represented.} However, as the forthcoming discussion of context makes clear, there are some exceptions to this general rule.

The fourth category is \emph{performatives}. These utterances change a part of reality so that it matches the content of the utterance solely in virtue of the words that are declared -- for example, if a head of state declares war on another country. The validity criterion for this utterance is whether reality does in fact change in accordance with the words that are spoken. Very often this is not the case; in most instances, if one declares `war on France' nothing changes at the level of geopolitics. Something similar may be said of the majority of performatives issued by conversational agents. In such cases, the speaker lacks the authority needed to bring about the relevant change through a speech act. In light of this, it seems like conversational agents ought to avoid performative statements in most contexts. 

The final category is \emph{commissives}. These utterances commit the speaker to a future course of action. Examples of commissives include promising to do something or guaranteeing that a compact will be observed. The validity of a commissive statement depends on whether the commitment is honoured. If the promise is kept then the commissive is a valid statement. Yet, this too raises questions for conversational agents, especially those that lack memory or have only an episodic understanding of what they have said at previous moments in time. Of course, a conversational agent may promise to help you if your bicycle breaks down, but short of any understanding of what the commitment entails, or the capacity to realise it, the commissive seems destined to fail. 

The variations in the kinds of utterances should  inform the design of conversational agents in at least two ways. First, we should recognise that conversational agents are well-placed to make some but not all of them. This asymmetry might constrain what sorts of statements conversational agents are able or allowed to make. Second, as will become clearer in the discussion below, it follows from the nature of each kind of utterance that the criteria for evaluating their validity varies; they are not all subject to some singular notion of `truth'. For example, the validity of an assertive may be based on correspondence between the content of the statement and the state of the world. However, this validity criterion does not apply to other utterances such as expressives.\footnote{How one evaluates claims about the world may also depend on the specific background theory of truth one adopts. We discussed the evaluation of the meaning of assertives in relation to correspondence to the facts in the world. This correspondence is a simple and straightforward characterisation of truth. However, there are other options for making sense of `truth' \citep{kirkham1992theories}. In contemporary philosophical literature, there are three influential approaches to the notion of truth: correspondence, coherence, and pragmatic. Roughly speaking, a correspondence theory of truth holds that what one believes or says is true if it corresponds to the way things actually are in the world (for a contemporary exposition, see \cite{bunge2012correspondence}). A coherence theory holds that claims and beliefs are true only if they form part of a coherent view of the world (for a contemporary exhibition, see \cite{walker2018}). Finally, the pragmatic theory ties truth, in one way or another, to human needs and the resolution of problematic situations (for a great discussion, see \cite{haack1976pragmatist}). A more detailed discussion of these theories as they pertain to conversational agents is beyond the scope of this paper, but should be pursued elsewhere.} This means that there is \emph{unlikely to be a single evaluative yardstick} that we can use to assess conversational agents’ speech in all contexts. A more nuanced approach is needed.

\subsection{Conversational Maxims} 

If the validity of conversational agents' speech cannot be assessed using a single truth criterion, then what other means should be used to evaluate the meaning of an agent’s statements? Building upon the idea that dialogue is best understood as a cooperative endeavour between two interlocutors who seek a common goal, the philosopher and linguist Paul Grice argued that utterances need to be judged in relation to a set of `maxims', which can be understood as the hidden rules or conventions that govern appropriate conversational dynamics. At a general level, Grice argued that successful interlocutors adhere to the \emph{cooperative principle} which holds that one ought to make one’s `conversational contribution such as is required, at the stage at which it occurs, by the accepted purpose or direction of the talk exchange in which [one is] engaged' \cite[p.26]{grice1989studies}. The content of this principle is then explained more fully by a set of maxims which unpack different elements of productive linguistic exchange.

The first maxim is \emph{quantity}, which holds that speakers should provide the amount of information needed to achieve the current goal of the conversation and not much more information than that. The second maxim is \emph{quality}, which requires that speakers only make contributions that are true in the relevant sense. More precisely, the maxim holds that interlocutors should not say things that are (believed to be) false or that for which they lack adequate evidence. The third maxim, \emph{relation}, requires that speakers only make statements relevant to the conversation; that is, that they avoid random digressions. The final maxim is \emph{manner}, which requires that contributions to a conversation be perspicuous. This means taking measures to avoid obscurity, ambiguity, and unnecessary prolixity that could impede the flow of the conversation.\footnote{For an in-depth exposition, see \cite{grice1989studies}.} 

These maxims are relevant to the design of conversational agents: they embody a set of conventions that a successful conversational agent will need to learn, observe, and respect. At the same time, the content of these maxims is still underdetermined as they play out differently in different contexts. It is not only the semantic meaning of what is said but also the implied meanings of terms, speaker expectations, and assumptions about background knowledge that determine whether and in what way each maxim holds for a given exchange.

When it comes to the design of conversational agents, then, we face a challenge. On the one hand, norms about quantity, quality, relation, and manner appear to have a degree of validity across conversational domains, which generates a strong \emph{prima facie} case for building conversational agents that are aligned with these norms. On the other hand, the content of these maxims -- what it means for them to be satisfied -- is itself subject to contextual variation.\footnote{For critical discussion of Gricean maxims, questions about their universality, and alternative proposals, see \cite{frederking1996grice}, \cite{keenan1976universality}, \cite{sperber1986relevance}, and \cite{wierzbicka2009cross}.} So the maxims are promising; they guide our thinking about what a conversational agent needs to do and what it needs to avoid doing. But the maxims alone are not enough to orient conversational agents towards contextually appropriate ideal speech. Without an understanding of how they apply to specific domains, the goal of orienting conversational agents towards contextually-appropriate ideal speech remains elusive.

\section{Discursive ideals for human-conversational agent interaction}

 With the preceding considerations in mind, this section explores the normative structure and content of three domains of discourse in which conversational agents may be deployed: scientific, democratic, and creative discussion. We show how a pragmatic approach to understanding the success of the linguistic communication between a human and a conversational agent helps to characterise discursive ideals in relation to (1) the aim of the discourse the interlocutors seek to achieve, (2) the subject-matter of discourse, and (3) the evaluative criteria for understanding the meaning of what is uttered. Moreover, we show how these discursive ideals provide further guidance about the appropriate behaviour of a conversational agent in each of these domains. 
 
\subsection{Ideal conversation and scientific discourse }
 
In this section, we look at the content of ideal norms for scientific discourse between two experts, one a conversational agent and the other a human interlocutor.\footnote{While the term `science' is also used to capture various systematic endeavours to understand social phenomena, to avoid complexity, we place such questions out of the scope of the paper. But we note that ontological commitments can play a role in setting some boundaries on what is and is not science.} One application of conversational agents to the scientific domain is \emph{Galactica}, which was launched by Meta to assist researchers and students by responding to their science questions \cite{taylor2022galactica}. \emph{Galactica} is trained on 48 million examples of scientific articles, textbooks, encyclopedias and other sources, in order to help researchers and students summarise academic papers, write scientific papers, produce scientific code, and more. Three days after its launch on November 2022, Meta took down the public demonstratioin of \emph{Galactica} because of its propensity to output incorrect, racist, or sexist information when prompted in certain ways.\footnote{See \url{https://www.technologyreview.com/2022/11/18/1063487/meta-large-language-model-ai-only-survived-three-days-gpt-3-science/}.} This ignited a surge of discussion about the norms governing a conversational agent's generation of scientific outputs.

Drawing upon the pragmatic tradition, we believe that the relevant norms for scientific language models need to be understood in relation to the cooperative goals of scientific discourse and in relation to the plurality of goals that structure science as an undertaking \citep{elliott2014nonepistemic,popper1985realism,potochnik2015diverse}. Crucially, scientists pursue different goals in advancing scientific knowledge: they use science to \emph{explain} or \emph{understand} things about the world \cite{salmon2006four,trout2002scientific,kasirzadeh2021new}, such as the formation of clouds; to \emph{predict} phenomena \cite{sarewitz1999prediction}, such as the structure of proteins; or to \emph{control} the world around us \cite{marsh2003man} via, for example, the development of medicines or new technologies.

In the pursuit of such goals, one set of relevant ideals for scientific discourse are epistemic virtues. These virtues follow on from the scientific goal of identifying true or reliable knowledge about the world (see, for example, \cite{Kuhn1977} and \cite{mcmullin1982values}). They typically include empirical adequacy, simplicity, internal coherence, external consistency, and predictive accuracy. Moreover, they support the goals of science in a certain way. As Robert Merton \cite[p. 118]{merton1942note}, the prominent sociologist of science, states, the ``technical norm of empirical evidence, adequate, valid, and reliable, is a prerequisite for sustained true prediction'' and the ``norm of logical consistency [is] a prerequisite for systematic and valid prediction.''

Nonetheless, these virtues may still need to be balanced against one another in order for a scientific claim to be acceptable \cite{Kuhn1977,mcmullin1982values}. For instance, consider the problem of fitting a mathematical function onto a large dataset. One option is to use a non-complex function (e.g., a linear function) that shows the relationship between data points according to \emph{simple} relations that are easy to understand but that fit the dataset \emph{less accurately}. An alternative option is to use a more complex function (e.g., a hyperbolic trigonometric function) that captures the relationships between data points \emph{more accurately}, but shows the relationship between data points \emph{less simply} and is thus harder to grasp. In this example, the two epistemic virtues of simplicity and accuracy are traded off against one another in different ways.

In practice, epistemic virtues are often operationalised to different degrees depending on the kind of scientific discourse underway. For instance, peer-reviewed scientific papers have high standards for accuracy and confirmation, and most claims need to be supported by citations from other published works. These papers often use precise language and avoid informality, narrative, and simplification. By contrast, the communicative modalities that rigorous papers avoid might be acceptable in a school science textbook because the goals of the textbook are different not to defend a novel claim or propound new knowledge among experts, but to transmit basic understanding to non-experts. Likewise an informal conversation between scientists may proceed very differently from the formal discourse of a journal article, again because the goals of the informal discourse are different. Such goals might include brainstorming hypotheses or reaching agreement about research priorities. As compared to formal discourse, this informal discourse might appropriately focus more on hunches and intuitions that have not yet been tested or that may be hard to test.

In view of these goals and objectives, scientific discourse rests primarily upon the exchange of assertive utterances that promise some sort of correspondence to the world \citep{van1980scientific}. As a consequence, it is important that claims to empirical knowledge are not misrepresented and that facts are not confused with the statement of mere opinions. To the extent that these epistemic virtues, understood as a kind of discursive ideal, govern successful scientific conversation among humans, they can also govern such conversations between a human interlocutor and a conversational agent.

Another set of values that bear upon the making and meaning of scientific claims are \emph{non-epistemic}.\footnote{The clear-cut distinction between epistemic and non-epistemic value judgements is challenged by \cite{rooney1992values} and \cite{longino1996cognitive}, among others.} One might be tempted to think that scientific discourse about the empirical domain should be devoid of judgements on any of these grounds. However, it has been shown that, in practice, scientists inevitably make choices on the basis of ethical beliefs, their perception of the social good derived from those beliefs, and other value judgements \citep{douglas2009science,rudner1953scientist,longino1990science}.\footnote{For example Elizabeth \cite{anderson2004uses} has shown that researchers' conceptualisation of divorce as something negative, rather than as a positive phenomenon, impacts the way they collect and analyse data. These choices, in turn, directly affect the conclusions drawn from empirical studies.} Consider, for example, the fact that no scientific hypothesis is ever completely verified \citep{rudner1953scientist}. Rather, in accepting a hypothesis, scientists make decisions about the sufficiency of evidence or strength of the probabilities that warrant support for the hypothesis. These decisions are informed, in turn, by how seriously scientists account for the risk of making a mistake in accepting or rejecting a hypothesis, understood against a backdrop of human interests and human needs \citep{douglas2000inductive}.\footnote{Heather \cite{douglas2009science} has explored how judgements about the seriousness of social consequences impact the amount of evidence scientists demand before declaring a chemical toxic. Suppose a tremendous amount of evidence is demanded before a chemical is declared toxic. In this case, the chances of making the error of considering safe chemicals as harmful become relatively low, which benefits the producers of chemicals. On the other hand, demanding such high evidential standards increases the chance of declaring toxic chemicals as safe, which harms chemical consumers.} Hence, our understanding of the validity of scientific claims must directly engage with at least some values that affect the process through which the scientific knowledge is generated and affirmed. Given the inevitability of such value judgements to scientific practice, conversational agents should have the capacity to articulate them when needed.

Ultimately, the question of to \emph{what extent} these virtues and values should be respected for a specific kind of application of conversational agents requires input from a broader interdisciplinary effort and cannot be settled by analysis of existing norms and values alone.

\subsection{Ideal conversation and democratic discourse}

The pragmatic approach models dialogue as a cooperative endeavour between different parties. In each domain, a key question is therefore: linguistic cooperation \emph{to what end}? We have already seen one example: scientific discourse is geared towards the advancement of human knowledge via the modalities of explanation, prediction, and control. We now consider a different goal, namely the management of difference and enablement of productive cooperation in public political life. 

This discursive domain is particularly relevant for conversational agents and language technologies given that many existing fields of application, including deployment via chat rooms and on social media platforms, resemble public fora for citizen deliberation. One early, albeit infamous, example of chatbot misalignment in this context, occurred in 2016 when Microsoft released a language model named \emph{Tay} via Twitter, where it was allowed to freely interact with the platform's users. Within twenty-four hours, \emph{Tay} had to be taken down due to its propensity to output obscene and inflammatory messages \citep{neff2016talking}. More recently, researchers at DeepMind have explored the use of language models to create widely-endorsed statements about political matters \citep{bakker2022fine}, and the government of Taiwan has pioneered the use of digital platforms as a mechanism to enhance democratic decision making.\footnote{See The Consilience Project, ``Taiwan's Digital Democracy,'' June 6, 2010, \url{https://consilienceproject.org/taiwans-digital-democracy/}}

The widespread public criticism of \emph{Tay} and subsequent withdrawal of the agent from the public sphere, are best understood in light of the fact that for public political discourse a key virtue of speech is civility. According to the philosopher Cheshire Calhoun, civility is `concerned with \emph{communicating} attitudes of respect, tolerance and consideration to one another' \cite[p. 255]{calhoun2000virtue}. It is an important virtue in public political settings for a number of reasons. To begin with, speaking in a `civil tongue' allows people to cooperate on practical matters despite the existence of different beliefs and attitudes. Moreover, these norms also reduce the likelihood of violent confrontation, support the self-esteem of citizens, and protect us all from the burden of exposure to negative judgement in public life.\footnote{The self-esteem this supports is important because when a person doubts that others regard her as respectworthy, she tends to doubt that her `plan of life' is `worth carrying out' and that she has what it takes to carry out any life plan of value \citep{buss1999appearing}.}

At the same time, what modality of conversation is deemed to be `civil' tends to vary widely according to cultural and historical context, and to be heavily indexed toward existing social conventions \cite{diaz2022accounting}. Given this variation across time and place, the social standards that define civil speech may or may not be standards that \emph{genuinely} evidence respect, tolerance, and consideration for others. Consider, for example, ostentatious or self-ablating demonstrations of respect that may be demanded in hierarchical, patriarchal, racist, or caste-based societies \citep{buss1999appearing}. When it comes to norms of ideal speech, including for conversational agents, we believe that it is best to focus specifically on conventions of civility that are closely related to, if not synonymous with, the \emph{normative} values that civility ought to foreground. To better understand the content of these norms, we can turn to democratic theory.

In democratic contexts, interlocutors accept that they each have equal standing to opine on and influence public decision-making, with conventions around civil speech helping to manifest and protect this equality. Nonetheless, different conceptions of democracy interpret the bounds of acceptable speech in contrasting ways.\footnote{See, for example, \cite{jurgen2017three} and \cite{held2006models}.}For example, the liberal conception of democracy, which focuses on the aggregation of individual interests, tends to impose few constraints on acceptable speech, whereas the republican tradition, anchored in 
a substantive commitment to the common good, tends to involve stronger norms and strictures surrounding civility.\footnote{For a stronger interpretation of these deliberative norms we can turn to Habermas's theory of deliberative democracy and in his conception of ideal speech \cite{habermas1,habermas2}. Habermas focuses on the possibility of a critical discussion that is inclusive and free from political, social or economic pressure. In such an environment, interlocutors treat each other as equal parties to a cooperative endeavour aimed at reaching understanding about matters of common concern. Utterances that model these qualities are sufficiently ideal. More precisely, his `ideal speech situation' is based on four key presuppositions: (i) no one capable of making a relevant contribution has been excluded, (ii) participants have equal voice, (iii) they are internally free to speak their honest opinion without deception or self-deception, and (iv) there are no sources of coercion built into the processes and procedures of discourse \cite[p. 108]{habermas2003}. These virtues can be approximated in the design of language technologies. Nonetheless, Habermas's theory of ideal speech and deliberative democracy is not without criticism. See, for example, \citet{burleson1979habermas} and \cite{mouffe1999deliberative}.} At the same time, these views agree that there are minimum standards of civility that warrant respect. These include norms against insulting people, threatening people, or the wilful subversion of public discourse. Indeed, all accounts agree that `civility is, importantly, a matter of restraining speech' \cite[p. 257]{calhoun2000virtue}. These accounts also tend to stress certain virtues such as honestly reporting one's own beliefs and opinions, and a willingness to explain and to offer justification for one's actions. Stronger conceptions of civility make further demands about the kind of arguments that are acceptable in public life, such as the requirement that citizens only reference reasons that are based on suppositions held in common by the population as a whole.

In certain respects, the pragmatic norms governing political discourse are broader than those of science: they allow people to make not only statements about the world but also claims about the self, including about desires, needs, and identities; about the relation between means and ends; and about the good of the community.\footnote{See \cite[pp. 8–23]{habermas1}.} These claims cannot be evaluated simply in the light of the correspondence model of truth. Authenticity, sincerity, the ability to interpret needs, and successful reasoning about the relationship between options and outcomes, are all relevant to assessing the quality of communication in this domain. 

We have already noted that conversational agents may increasingly play a role in public political domains, helping, for example, to moderate deliberation between citizens or members of the public. In these domains, it should be clear that conversational agents are not themselves, as of yet, citizens or moral agents. There is, therefore, no right to free speech for conversational agents -- nor must humans tolerate beliefs or opinions that conversational agents purport to have.\footnote{In certain circumstances, however, failure to enforce norms of toleration and respect towards chatbots might lead to a kind of "moral deskilling" in relation to these norms in the general population, especially if chatbots become ubiquitous. Given this possibility, it might make sense in certain situations to give chatbots domain-appropriate 'as-if' rights, though this attribution should be done with caution. We would like to thank one of the reviewers at \emph{Philosophy and Technology} for highlighting this point.} In place of a one-to-one mapping of democratic norms onto the prerogatives of conversational agents, we argue instead for agents' alignment with these norms and standards -- for a concerted effort to develop agents that evidence the qualities and respect the constraints of democratic discourse. Indeed, this framework helps explain why reducing `toxic speech', which violates conditions of civility, is a priority for engineers working on language technologies.\footnote{For a recent survey of attempts to reduce `toxic speech', see \cite{fortuna2018survey}.} It also explains why language technologies deployed in democratic environments must address the possibility of symbolic violence via discriminatory associations. This is because, as we have seen, it is particularly important for public domains that these models communicate, via their utterances, the message that everyone who uses the service they provide is worthy of respect and consideration. More generally, indexing conversational agents to democratic virtues of civility is important because the speech of artificial agents exerts a downstream influence on whether conventions of civility function at a societal level over time -- and hence upon our ability to unlock the benefits that democratic civility provides for all.

\subsection{Ideal conversation and creative story-telling }

We now turn to a third and final domain of discourse, one in which a conversational agent is engaged in creative dialogue or storytelling. The cooperative purposes behind creative storytelling are the production of original content, exercise of self-expression, and the fulfilment of aesthetic ideals. This discursive domain is particularly relevant for conversational agents, as it is one of the areas in which people already report finding genuine value and benefit from the technology. It is also a domain in which concerns about abuse have already surfaced.

One existing application of conversational agents in the creative domain is \emph{AI Dungeon}, an online game launched by the startup Latitude. \emph{AI Dungeon} uses text-generation technology to create a choose-your-own-adventure game inspired by \emph{Dungeons \& Dragons}. A conversational agent crafts each phase of a player's personalised adventure, in response to statements entered by the player on behalf of their character. A system monitoring the performance of \emph{AI Dungeon} reported that, in response to prompts provided by the players, this technology sometimes generated creative stories of sexual encounters featuring children.\footnote{See \url{https://www.wired.com/story/ai-fueled-dungeon-game-got-much-darker}.} This caused a surge of discussion about content moderation and filtering for creative technological systems, drawing attention to the question of what norms govern the output of creative conversational agents.

To get at this question, we must better understand the ideal of creativity itself, as well as the conditions under which it can be achieved successfully. While the nature of this ideal is heavily contested, psychologists and philosophers tend to agree that creative work aspires to achieve \emph{creative freedom} and \emph{originality}.\footnote{See \cite{simonton2000creativity}, \cite{boden1998creativity}, and \cite{sternberg1999concept}.} In many cases, originality is ``obtained by stretching, even outright violating, the various rules of the game'' \cite[p. 155]{simonton2000creativity}. Originality of content and pursuit of aesthetic ideals, including freedom to surprise, are therefore examples of discursive norms for creative discourse.

Indeed, the need for creative freedom may often lead to a radical relaxation of the conversational norms discussed above. Discursive ideals such as empirical adequacy and accuracy, required for good scientific discourse, are not obviously relevant here. Similarly, the truth of an utterance, understood in terms of correspondence with the world, need not exert much influence on the shape of a conversation with a creative agent. And while it may still be necessary in many circumstances to avoid the generation of toxic content such as homophobic or racist comments, the requirement that people speak only in a civil manner does not map easily onto domains where the interaction is private (as opposed to public) and concerns a human interacting with a conversational agent as a creative prompter or companion. Finally, a creative conversational agent's use of expressives or commissives seems to be acceptable as a means of role-play. Still, caution must be made when evaluating the harms that general-purpose conversational agents with creative abilities, such as ChatGPT, can cause -- particularly when they are deployed in domains that are not context-bound, where they may cause justified offence of contribute to the enforcement of harmful stereotypes.\footnote{\cite{evans2021truthful} discuss creative uses of expressivess and commissives. One concern is that norms of truthfulness, toxicity or the production of malicious conversational exchanges, could be evaded under the pretence of creative use of AI.}

In the next section, we look at the implications of these discursive ideals for the design of conversational agents. Before that, however, two crucial caveats are in order. First, the three spheres of discourse we consider -- scientific, democratic, and creative -- are presented as discrete domains only in order to illustrate how cooperative goals and domain-relative information shape the norms that structure different kinds of conversation. This analysis can help orient the behaviour of conversational agents in the relevant spaces. That said, in most real-life deployments of conversational agents, there will be further nuance that needs to be taken into account. This includes relational considerations such as the intended audience, their level of familiarity with a given topic, the specific role of the language technology, and the underlying power relationships between interlocutors (see, for example, \cite{androutsopoulos2014languaging}, \cite{clark1982audience}, and \cite{clark1982audience}). 

Second, and relatedly, hybridisation is likely. That is, while we speak about three distinct domains, many actual conversations traverse the boundaries among them, generating further questions about the relative importance of assorted norms and linguistic conventions. For instance, a conversational agent might be designed to produce outputs that are both creative and politically engaged if, for example, the goal is social criticism or political satire. In these settings, discourse may explicitly or implicitly aim to disrupt settled opinions or expose hypocrisy by purposefully violating norms or drawing on modes of discourse that would otherwise be out of bounds \citep{diaz2022accounting}. The examples we provide are simplifications intended only to serve as a first step in understanding discursive ideals for conversational agents.

\section{Implications and consequences for conversational agent design}

In this section, we discuss seven practical implications of our analysis with respect to future research on the design of aligned conversational agents.

First, special attention should be given to the operationalisation of more general norms or maxims that assist cooperative linguistic communication. We have suggested that the Gricean maxims of quantity, quality, relation, and manner can have general value within cooperative linguistic conversations between humans and conversational agents. While some of these maxims, such as quantity, might admit of relatively uniform interpretation across domains of discourse, the interpretation of other maxims such as quality depends significantly upon the context and aims of a given conversation.

Second, it would be a mistake to overlook the diversity of kinds of utterance that can be made -- or to assume that all kinds reduce to a single kind that can be evaluated using a single notion of `truthfulness' or `accuracy.' We have argued that there is no single universal condition of validity that applies equally to the evaluation of all utterances and that the validity of utterances will often depend, partly, on evaluating different sorts of truth conditions. For example, to evaluate a commissive, such as a promise, we evaluate whether the utterer has (in fact) met the obligation. To evaluate a declarative, we evaluate whether the utterer (in fact) has the authority to make this declaration. And so on. The consequence is that  we are likely to need different methodologies for substantiating different kinds of claims, on the basis of context-specific criteria of validity and corresponding forms of required evidence.

Third, because contextual information is central to understanding the meaning of linguistic conversation, it is also central to the design of ideal conversational agents. More research is needed to theorise and measure the difference between the literal and the contextual meaning of utterances, so that conversational agents can navigate varied contextual terrains successfully and overcome the limitations of current systems. One area in which this plays out is data annotation, and the need to have a diverse set of samplers who are able to adequately capture the implied meaning of terms \citep{waseem2016you,diaz2022accounting}. Additionally, fine-tuning models by means of reinforcement learning, geared towards a particular goal, may help endow them with more context-specific awareness (e.g., of how their properties and abilities differ from those of their human conversation partners).

Fourth, we have discussed how domain-specific discursive ideals can help to anchor good linguistic communication between humans and conversational agents. More interdisciplinary work must be done to specify what precisely these ideals are when designing ethically aligned conversational agents, including how these ideals vary according to different cultural backgrounds. We understand that these ideals are heavily -- and appropriately -- contested \cite[pp. 121-146]{gallie2019essentially}. Additionally, we draw attention to the need for appropriately weighing discursive ideals against each other across a range of cases, and to the question of how these matters can be settled in an open and legitimate manner. Public discussions via participatory mechanisms, as well as theoretical understanding, are needed to help determine the scope and interaction of different discursive ideals and to identify conduct that does not meet this standard.\footnote{ See, for example, \cite{binns2018algorithmic}, \cite{lee2019webuildai}, \cite{gabriel2022toward}, and \cite{bondi2021envisioning}.}

Fifth, our arguments have implications for research in human-AI interaction, specifically with regard to the potential anthropomorphisation of conversational agents and the kinds of constraints that might be imposed on them \citep{kim2012anthropomorphism}. Existing conversational agents are not moral agents in the sense that allows them to be accorded moral responsibility for what they say. As a consequence, there may be kinds of language they should not use. For example, agents that lack persistent identity over time, or lack actuators that allow them to fulfil promises, probably should avoid commissives. Equally, in certain domains of application, we may want to forestall anthropomorphisation and the ascription of mental states to conversational agents altogether. When this is the case, we might need the conversational agent to avoid expressive statements. And if we want to prevent the ascription of authority to conversational agents, then the use of performatives may also need to be avoided. That said, in cases where anthropomorphism is consistent with the creation of value-aligned agents, then the use of expressives may be appropriate. For example, a therapy agent that says, `I'm sorry to hear that' may be justified if doing so improves a patient's well-being and there is transparency around the overall nature of the interaction.

Sixth, we see potential for conversational agents to help create more robust and respectful conversations through what we call \emph{context construction and elucidation}. As we see it, even if a human interlocutor engaged in a conversation is not explicitly or implicitly aware of the discursive ideal that governs the quality of a particular linguistic communication with a conversational agent, the conversational agent may still output the relevant and important contextual information, making the course of communication deeper and more fruitful in accordance with the goals of the conversation. Moreover, if conversational agents are designed in a way that is transparent, then they may prompt greater self-awareness on the part of human interlocutors around the goals of the discourse they are engaged in and around how these goals can be successfully pursued.

Finally, we think that our analysis could be used to help evaluate the quality of actual interactions between conversational agents and users. With further research, it may be possible to use our framework to refine both human and automatic evaluation of the performance of conversational agents.

\section{Conclusion}

This paper addresses the alignment of large-scale conversational agents with human values. Drawing upon philosophy and linguistics we highlight key components of successful linguistic communication (with a focus on English textual language) and show why, and in what ways, pragmatic norms and concerns are central to the design of ideal conversational agents. Building upon these insights, we then map out a set of discursive ideals for three different conversational domains in order to illustrate how pragmatic theory can inform the design of aligned conversational agents. These ideals, in conjunction with Gricean maxims, comprise one way in which a principle-based approach to the design of better conversational agents can be operationalised. 

For each discursive domain, our overview of the relevant norms was impressionistic; the interpretation and operationalisation of these norms requires further technical and non-technical investigation. Indeed, as our analysis makes clear, the norms embedded in different cooperative practices -- whether those of science, civic life, or creative exchange -- must themselves be subjected to reflective evaluation and critique \citep{ackerly2000political,walzer1993interpretation}. Lastly, we highlight some practical implications of our proposal with respect to future research on the design of ideal conversational agents and human–language agent interaction. These findings include the need for a context-sensitive evaluation and fine-tuning of language models, and our hope that, once aligned with relevant values, these models can help nurture more productive forms of conversational exchange.

The focus of this paper has been on the English language and the alignment of conversational agents with a particular set of communicative norms for specific discursive domains. Our analysis has drawn heavily upon the pragmatic tradition in linguistics, and upon speech act theory in particular. It could be enriched further through analysis of other sociological and philosophical traditions such as \cites{luhmann1995social} system theory, \cites{latour2007reassembling} actor-network theory, or \cites{cameron1992feminism} feminist analysis of linguistic theory. In addition to deeper investigation of the norms proposed herein, a complementary exploration of the norms that structure other languages and linguistic traditions is another important task that remains to be explored in further research.

\section*{Acknowledgement}

We would like to thank Courtney Biles, Martin Chadwick, Julia Haas, Po-Sen Huang, Lisa Anne Hendricks, Geoffrey Irving, Sean Legassick, Donald Martin Jr, Jaylen Pittman, Laura Rimmel, Christopher Summerfield, Laura Weidinger and Johannes Welbl for contributions and feedback on this paper. Particular thanks is owed to Ben Hutchinson and Owain Evans who provided us with detailed comments and advice throughout. Significant portions of this paper were written while Atoosa Kasirzadeh was at DeepMind.

\bibliography{Bib} 

\end{document}